\documentclass[prl,aps,superscriptaddress,amssymb,twocolumn]{revtex4-1}
\usepackage{graphicx}
\usepackage{amsmath}
\usepackage{latexsym}
\usepackage{dcolumn}
\usepackage{bm}
\usepackage{float}
\usepackage{graphicx,here}

\emergencystretch=\maxdimen
\begin{document}
\title{Superstructure-induced splitting of Dirac cones in silicene}

\author{Baojie Feng}
\thanks{Email: bjfeng@iphy.ac.cn}
\affiliation{Institute of Physics, Chinese Academy of Sciences, Beijing 100190, China}
\author{Hui Zhou}
\affiliation{Institute of Physics, Chinese Academy of Sciences, Beijing 100190, China}
\affiliation{School of Physical Sciences, University of Chinese Academy of Sciences, Beijing 100049, China}
\author{Ya Feng}
\affiliation{Ningbo Institute of Materials Technology and Engineering, Chinese Academy of Sciences, Ningbo 315201, China}
\affiliation{Hiroshima Synchrotron Radiation Center, Hiroshima University, 2-313 Kagamiyama, Higashi-Hiroshima 739-0046, Japan}
\author{Hang Liu}
\affiliation{Institute of Physics, Chinese Academy of Sciences, Beijing 100190, China}
\affiliation{School of Physical Sciences, University of Chinese Academy of Sciences, Beijing 100049, China}
\author{Shaolong He}
\affiliation{Ningbo Institute of Materials Technology and Engineering, Chinese Academy of Sciences, Ningbo 315201, China}
\author{Iwao Matsuda}
\affiliation{Institute for Solid State Physics, The University of Tokyo, Kashiwa, Chiba 277-8581, Japan}
\author{Lan Chen}
\affiliation{Institute of Physics, Chinese Academy of Sciences, Beijing 100190, China}
\affiliation{School of Physical Sciences, University of Chinese Academy of Sciences, Beijing 100049, China}
\author{Eike F. Schwier}
\affiliation{Hiroshima Synchrotron Radiation Center, Hiroshima University, 2-313 Kagamiyama, Higashi-Hiroshima 739-0046, Japan}
\author{Kenya Shimada}
\affiliation{Hiroshima Synchrotron Radiation Center, Hiroshima University, 2-313 Kagamiyama, Higashi-Hiroshima 739-0046, Japan}
\author{Sheng Meng}
\thanks{Email: smeng@iphy.ac.cn}
\affiliation{Institute of Physics, Chinese Academy of Sciences, Beijing 100190, China}
\author{Kehui Wu}
\thanks{Email: khwu@iphy.ac.cn}
\affiliation{Institute of Physics, Chinese Academy of Sciences, Beijing 100190, China}
\affiliation{School of Physical Sciences, University of Chinese Academy of Sciences, Beijing 100049, China}

\date{\today}

\begin{abstract}

Atomic scale engineering of two-dimensional materials could create devices with rich physical and chemical properties. External periodic potentials can enable the manipulation of the electronic band structures of materials. A prototypical system is 3$\times$3-silicene/Ag(111), which has substrate-induced periodic modulations. Recent angle-resolved photoemission spectroscopy measurements revealed six Dirac cone pairs at the Brillouin zone boundary of Ag(111), but their origin remains unclear [Proc. Natl. Acad. Sci. USA 113, 14656 (2016)]. We used linear dichroism angle-resolved photoemission spectroscopy, the tight-binding model, and first-principles calculations to reveal that these Dirac cones mainly derive from the original cones at the K (K$^{\prime}$) points of free-standing silicene. The Dirac cones of free-standing silicene are split by external periodic potentials that originate from the substrate-overlayer interaction. Our results not only confirm the origin of the Dirac cones in the 3$\times$3-silicene/Ag(111) system, but also provide a powerful route to manipulate the electronic structures of two-dimensional materials.

\end{abstract}

\maketitle

Two-dimensional (2D) honeycomb lattices of group IV elements, such as graphene, host Dirac cones at the K (K$^{\prime}$) points of the Brillouin zone (BZ)\cite{Neto2009,KotovVN2012,ZhuangJ2015,ZhaoJ2016}. Manipulation of the physical properties of Dirac fermions provides a powerful route to realizing exotic quantum devices. For example, employment of symmetry-breaking potentials could renormalize the Fermi velocity of the Dirac bands\cite{ParkCH2008,RusponiS2010,ZouQ2013}, replicate the Dirac cones\cite{ParkCH2008',BarbierM2010}, or gap out the Dirac points\cite{PletikosicI2009,LiuW2009,DvorakM2013,Song2013}, offering great opportunities to manipulate their transport properties. In the three-dimensional counterpart of graphene, namely Dirac semimetals, breaking the inversion or time-reversal symmetry could lead to the momentum-splitting of a Dirac cone into a pair of Weyl cones, giving rise to rich physical properties such as chiral anomaly or extremely large magnetoresistance\cite{ArmitageNP2018,YanB2017}. In the 2D limit, however, momentum-splitting of the Dirac cones remains a challenging task in a simple honeycomb lattice.

A promising platform for studying the effects of external potentials on Dirac fermions is silicene, which can form various short-range superstructures on Ag(111)\cite{VogtP2012,FengB2012,MengL2013}. In particular, the 3$\times$3 superstructure of silicene (or 4$\times$4 superstructure of Ag(111)) has been intensively studied. As is the case with graphene, theoretical calculations have predicted that free-standing silicene hosts a Dirac cone at each K (K$^{\prime}$) point\cite{CahangirovS2009,LiuCC2011}. In 3$\times$3-silicene/Ag(111), however, early angle-resolved photoemission spectroscopy (ARPES) measurements and theoretical calculations showed that the Dirac cones of free-standing silicene are destroyed because of the strong band hybridization between silicene and Ag(111)\cite{VogtP2012,TsoutsouD2013,MahathaSK2014,XuX2014,FengY2016,Sheverdyaeva2017}. Recently, high-resolution ARPES experiments revealed the existence of six pairs of Dirac cones at the Brillouin zone (BZ) boundary of Ag(111), but these Dirac cones originate from the substrate-overlayer interaction instead of from the free-standing silicene\cite{FengY2016}. To date, the physics on the origin of the Dirac cone pairs is still unclear despite the recent success of band unfolding analysis based on density functional theory (DFT)\cite{IwataJI2017,LianC2017}.

In this Letter, we will show that the Dirac cone pairs in 3$\times$3-silicene/Ag(111) derive mainly from the original Dirac cones of free-standing silicene. The Dirac cones of free-standing silicene are split by periodic perturbations that are related to the 3$\times$3 superstructure. These results were supported by our linear dichroism ARPES measurements, tight-binding analysis, and first-principles calculations. Our results not only settle the long-debated question on the existence of Dirac cones in silicene/Ag(111), but also provide a powerful route to tailor the physical properties of Dirac fermions in a 2D material.

\begin{figure*}[htb]
\includegraphics[width=16cm]{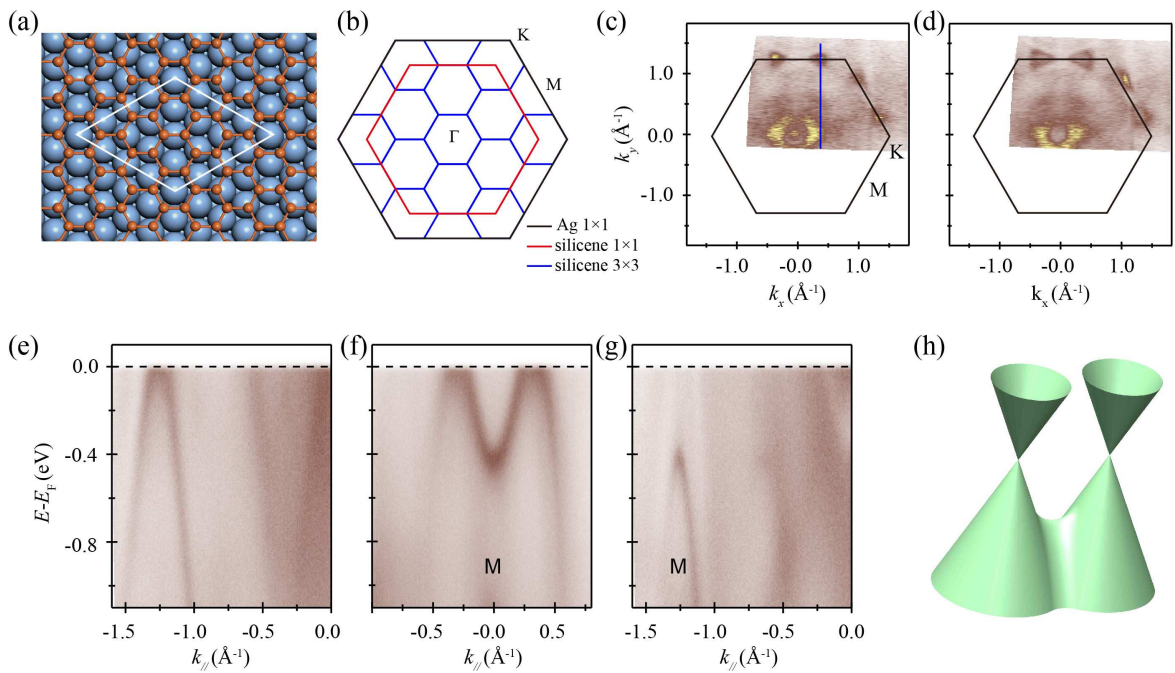}
\caption{(a) (Color online) Atomic structure of 3$\times$3 silicene on Ag(111). Orange and blue balls represent Si and Ag atoms, respectively. The white rhombus indicates a unit cell of the 3$\times$3 superstructure. (b) BZs of silicene (red lines), Ag(111) (black lines), and the superstructure (blue lines). (c) Photoemission intensity at the Fermi level as measured with 50 eV $p$-polarized photons. The blue line indicates the momentum cut along which the ARPES intensity plots in (e) were taken. (d) The same plot as in (c) but at a binding energy of 0.2 eV. (e-g) ARPES intensity plot along the blue line in (c), along the $\rm\bar{K}$-$\rm\bar{M}$-$\rm\bar{K}$ direction, and along the $\bar{\Gamma}$-$\rm\bar{M}$ direction, respectively. (h) Schematic drawing of a pair of Dirac cones.}
\end{figure*}

Both the sample preparation and ARPES measurements were performed at the linear undulator beamline BL-1 of the Hiroshima Synchrotron Radiation Center. Monolayer silicene was prepared by evaporating silicon from a silicon wafer onto a Ag(111) substrate. Clean Ag(111) surfaces were obtained by repeated sputtering and annealing cycles. During the growth of silicene, the substrate temperature was kept at approximately 460 K. After preparation, the samples were quickly transferred to the ARPES chamber without breaking the vacuum. The linear polarization of the incident light could be switched between {\it s} and {\it p} by rotating the whole ARPES system; this ensured pure {\it s} and {\it p} polarization without cross contamination as is the case with undulator-based polarization switching\cite{IwasawaH2017}. The overall energy and angular resolutions were approximately 15 meV and 0.1$^{\circ}$, respectively. The temperature of the sample during the measurements was kept at 40 K.

First-principles calculations based on DFT were performed with the Vienna Ab-initio Simulation Package (VASP)\cite{KresseG1993,KresseG1996}. The projector-augmented wave pseudopotential and Perdew-Burke-Ernzerhof exchange-correlation functional were used\cite{BlochlPE1994}. The energy cutoff of the plane-wave basis was set as 250 eV, and the vacuum space was set to be larger than 15{\AA}. For the 3$\times$3-silicene/Ag(111) sample, we constructed the structure model by placing monolayer silicene on a four-layer Ag(111) slab. The first BZ was sampled according to the $\Gamma$ centered $k$-mesh. We used a $k$-mesh of 6$\times$6$\times$1 for structural optimization and 12$\times$12$\times$1 for the self-consistent calculations. The positions of the atoms were optimized until the convergence of the force on each atom was less than 0.005 eV/{\AA}. The convergence condition of the electronic self-consistent loop was 10$^{-5}$ eV.

Figure 1(a) shows the structure model of 3$\times$3-silicene on Ag(111). The white rhombus indicates a unit cell of the superstructure which includes a 3$\times$3 superlattice of silicene and a 4$\times$4 superlattice of Ag(111). From the Fermi surface in Fig. 1(c), we found that there exist a pair of Fermi pockets symmetric to the $\rm\bar{M}$ point of Ag(111). The size of the pockets increases at higher binding energies, as shown in Fig. 1(d). Figure 1(e) shows the band structure along the blue line in Fig. 1(c), which shows a linear dispersion. The band structure along the $\rm\bar{K}$-$\rm\bar{M}$-$\rm\bar{K}$ direction also shows a linear dispersion. These results indicate the existence of a pair of Dirac cones at the BZ boundary of Ag(111), which is consistent with previous ARPES results\cite{FengY2016}. It should be noted that our results were obtained with a different photon energy; this is reasonable because these bands originate from the surface layers and have a two-dimensional character. The Dirac points are located above the Fermi level, which was not accessible by our ARPES.

The emergence of Dirac cones at the BZ boundary of Ag(111) is peculiar since they are not located at high symmetry points of the BZ of either silicene or Ag(111). However, we realized that the bands along the $\rm\bar{K}$-$\rm\bar{M}$-$\rm\bar{K}$ direction exhibit a parabolic dispersion with the band bottom located at the $\rm\bar{M}$ point of Ag(111). Along the perpendicular direction, i.e., the $\bar{\Gamma}$-$\rm\bar{M}$ direction, we found a reversed parabolic dispersion with the band top located at the $\rm\bar{M}$ point. At the $\rm\bar{M}$ point, the band bottom along the $\rm\bar{K}$-$\rm\bar{M}$-$\rm\bar{K}$ direction and band top along the $\bar{\Gamma}$-$\rm\bar{M}$ direction are located at the same binding energy, i.e., approximately 0.4 eV below the Fermi level. These results indicate the existence of a saddle point between the pair of Dirac cones (as illustrated in Fig. 1(e)) and agree with previous ARPES results\cite{TsoutsouD2013,FengY2016}. The existence of a saddle point indicates that the pair of Dirac cones have a common origin and is reminiscent of Weyl cone pairs that are split from a Dirac cone after certain symmetry breaking. These results suggest the possibility that the pair of Dirac cones are split from one original cone at the $\rm\bar{M}$ point of Ag(111).

To confirm this, we first discuss the possible existence of such an original Dirac cone (before splitting) at the $\rm\bar{M}$ point of Ag(111). From the schematic drawing of the BZs, we can see that the K point of the silicene 1$\times$1 lattice is equivalent to the $\rm\bar{M}$ point of Ag(111), because both of them are located at the $\Gamma$ point of the superstructure (blue hexagons). In analogy to graphene, freestanding silicene naturally hosts a Dirac cone at each K (K$^{\prime}$) point; this Dirac cone will be folded to the $\rm\bar{M}$ point of Ag(111) in the presence of the superstructure. Another important fact is that the Dirac cones of a honeycomb lattice mainly derive from the $p_z$ orbitals of the outermost shell. We will show next that this can be experimentally confirmed by our ARPES measurements.

To disentangle the orbital composition of the Dirac cones, we performed linear dichroism ARPES. In our ARPES measurement setup, the $xz$ plane is the incident plane as well as the detection plane, as shown in Fig. 2(a). When the detection plane coincides with the mirror plane of the sample, there will be a selection rule in the photoemission process according to the electric dipole approximation: linearly polarized light tends to excite electronic states that have well-defined parities with respect to the mirror plane: the $p$-polarized light excites the even-parity states while the $s$-polarized light excites the odd-parity states\cite{Damascelli2003}. In silicene, the low-energy electronic states are dominated by the $p$ electrons, including the $p_x$, $p_y$, and $p_z$ orbitals. In particular, the $p_z$ orbitals always have an even parity when we rotate the sample around the $z$ axis, and thus can only be excited by $p$-polarized light. The photoemission intensity along the $\bar{\Gamma}$-$\rm\bar{D}$ direction measured with $p$ and $s$ polarized light is shown in Fig. 2(b). One can see that the linearly dispersing Dirac bands are only visible with $p$ polarized light, while they almost disappear with $s$ polarized light; this result agrees well with the fact that the Dirac cones mainly originate from the $p_z$ orbitals. It should be noted that the $\bar{\Gamma}$-$\rm\bar{D}$ direction is not on a mirror plane of the silicene/Ag(111) system, so the distinct polarization dependence of the Dirac cones along the $\bar{\Gamma}$-$\rm\bar{D}$ direction indicates that the Dirac cones are mainly derived from the $p_z$ orbitals, in agreement with the parity invariance of the $p_z$ orbitals under the rotation of the sample about the $z$ axis. These results further corroborate our speculation that the Dirac cones originate from the original ones of the free-standing silicene.

\begin{figure}[htb]
\includegraphics[width=8cm]{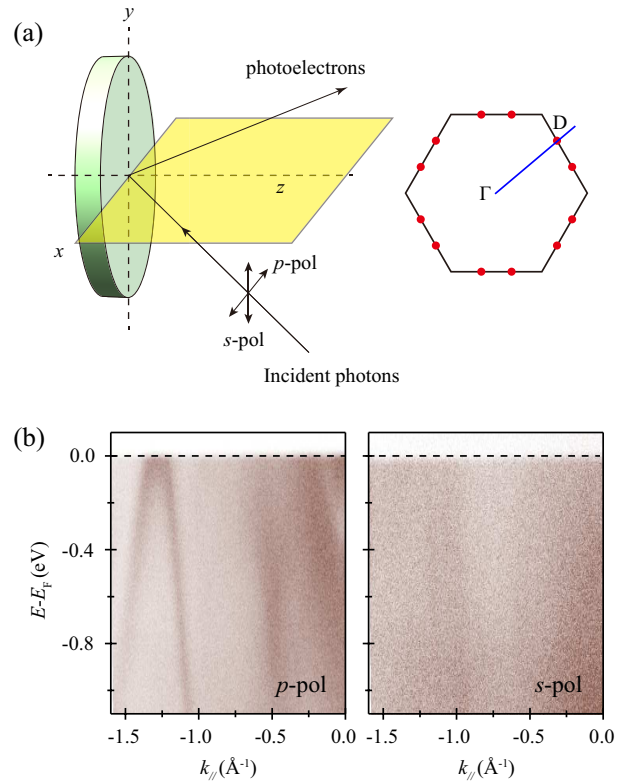}
\caption{(Color online) (a) Left panel: schematic illustration of the experimental geometries of the ARPES measurements. The incident light lies in the detection plane (colored yellow). The $s$ and $p$ polarization is defined as perpendicular and parallel to the detection plane, respectively. Right panel: schematic drawing of the BZ of Ag(111) and the positions of the Dirac cones. The blue line indicates the momentum cut that includes the $\Gamma$ point and Dirac point. (b) ARPES intensity plots along the blue line with $p$ and $s$ polarized light, respectively. The incident photon energy is 50 eV.}
\end{figure}

\begin{figure*}[htb]
\includegraphics[width=12cm]{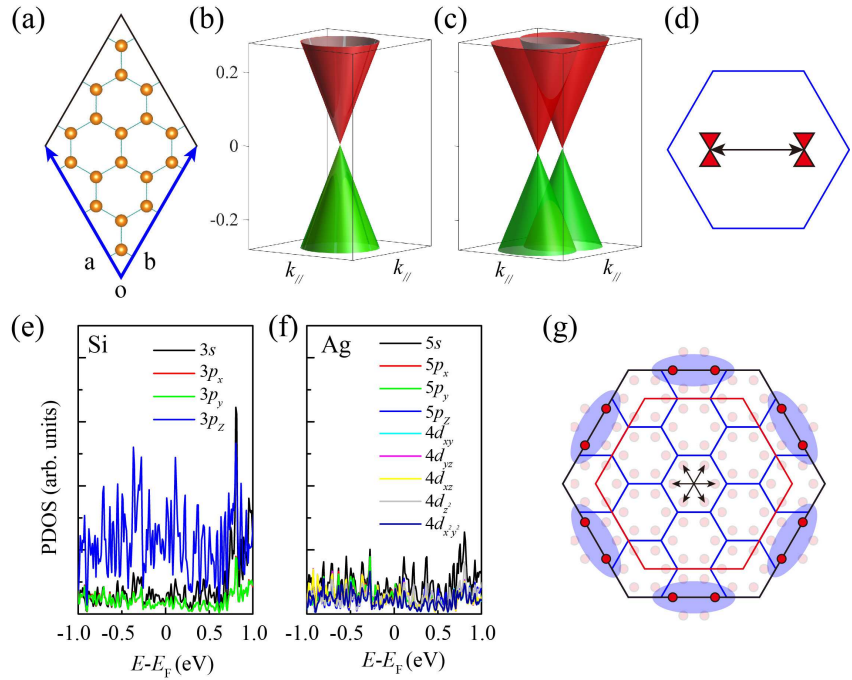}
\caption{(Color online) (a) Schematic illustration of the periodic potential in our tight-binding model. (b) and (c) Calculated band structures using the tight-binding model before (b) and after (c) applying the periodic potential. (d) Schematic drawing of the splitting direction of the Dirac cone. (e) and (f) Calculated partial density of states of Si and Ag atoms, respectively, using DFT. (g) Schematic drawing of the Dirac cone pairs in the BZs of silicene and Ag(111). The red dots represent the positions of the Dirac cones. The blue shaded ellipses highlight the Dirac cones that have been experimentally observed. The black double arrows indicate the splitting of the Dirac cones.}
\end{figure*}

Now that we have identified the origin of the Dirac cones in 3$\times$3-silicene/Ag(111), the next question is why the Dirac cones are split into pairs. In 3$\times$3-silicene/Ag(111), the Si atoms are slightly buckled, forming a 3$\times$3 superstructure. The buckling of Si is approximately 0.8 {\AA}\cite{FukayaY2013,CurcellaA2016}, which is close to the value of the low-buckled structure of free-standing silicene\cite{CahangirovS2009}. As a result, we conclude that the interaction between silicene and Ag(111) is not very strong. To capture the physical mechanism of the splitting, it is convenient to use a simple tight-binding model in which the effects of the substrates are treated as perturbations. We start with a flat silicene structure, and only the $p_z$ orbitals of Si are taken into account. The Hamiltonian of silicene takes the form:

\begin{equation}
\mathcal{H}=-t_{0}\sum_{ij}^{\text{n.n.}}c_{i}^{\dagger}c_{j}, \label{eqs3}%
\end{equation}

where $c_{i}^{\dagger}$ and $c_{j}$ are the creation and annihilation operator of Si $p_z$. In this case, there is a Dirac cone centered at each K (K$^{\prime}$) point of the BZ, as shown in Fig. 3(b). When silicene is placed on Ag(111), a 3$\times$3 superstructure will form, and, more importantly, the sublattice symmetry will be broken\cite{LinCL2013}. In our tight-binding model, we used an external perturbation to simulate the effects of the substrate. From this point of view, the superstructure has two prominent effects. First, the band structures will be folded into the BZs of the 3$\times$3 supercell because of the Umklapp scattering process, resulting in the appearance of a Dirac cone at the $\rm\bar{M}$ point of Ag(111), as discussed above. Second, it will expose a long-range periodic potential in silicene, which has not yet been taken into account. To simulate the effects of the periodic potential imposed by the substrate, we varied the on-site energy of each Si atom in the 3$\times$3 supercell according to the following equation:

\begin{equation}
V_{m,n}(x,y)=V_0cos(2\pi mx/L_a)cos(2\pi ny/L_b)
\end{equation}

where m and n are integers and $L_a$=$L_b$ ($L_a$ and $L_b$are the lattice constants of the supercell). Applying the periodic potential will break the sublattice symmetry and lead to a dramatic change in the Dirac cone. With proper parameters\cite{SM}, such as m=4 and n=9, the Dirac cone will split into a pair along the $\Gamma$-K direction of the 3$\times$3 supercell. These results could qualitatively explain the origin of Dirac cone pairs that are symmetric to the $\rm\bar{M}$ point of Ag(111).

A quantitative explanation of our experimental results requires DFT calculations that include the Ag(111) substrates. In fact, recent band-unfolding methods based on DFT have shown signatures of the Dirac cone pairs\cite{IwataJI2017,LianC2017}. Moreover, our calculated partial density of states near the Fermi level are dominated by the $p_z$ orbitals of Si, as shown in Fig. 3. These results agree well with our experimental results and support our tight-binding model that only consider the p$_z$ orbitals of Si. On the other hand, although the contributions from the Ag orbitals are smaller than those from Si $p_z$ orbitals, there are still considerable contributions from some orbitals of Ag, such as the $s$, $p_z$, and $d_{z^2}$ orbitals. These orbitals have out-of-plane components and are prone to hybridize with Si $p_z$ orbitals. This band hybridization, although weak, imposes a periodic potential in the silicene lattice and results in the splitting of the Dirac cone.

In our ARPES measurements, the Dirac cones are clearly visible at the BZ boundary of Ag(111), while their intensity is much weaker elsewhere, as schematically shown in Fig. 3(g). One possible reason is that the bulk $sp$ bands of Ag(111) are located at the BZ boundary. In the photoemission process, the emitted photoelectrons are compensated by the itinerant electrons of the Ag(111) substrate. Since the Shockley surface states of Ag(111) disappear after adsorption of silicene, only electrons from the bulk $sp$ bands are the possible electron source. The electrons from the substrate preferably compensate the nearby electronic states, which leads to an enhancement of the Dirac cones at the BZ boundary of Ag(111). Another possible reason is that there are photoemission matrix element effects, since the photoemission intensity is dependent on various factors, such as the photon energy, polarization, and incident angle.

In summary, we studied the electronic structures of 3$\times$3-silicene on Ag(111) using high-resolution ARPES, tight-binding analysis, and DFT calculations. We confirmed the existence of six pairs of Dirac cones at the BZ boundary of Ag(111), but these Dirac cones were only detectable with {\it p} polarized light, indicating that the Dirac bands are mainly derived from the $p_z$ orbitals of silicon. Our tight-binding analysis and DFT calculations revealed that these Dirac cones originate from the original Dirac cones of free-standing silicene. Our results settle the long-debated question on the existence of Dirac cones in the silicene/Ag(111) system, and, more importantly, provide a powerful route to tailor the physical properties of Dirac fermions in a honeycomb lattice.

\begin{acknowledgments}
{\center\bf Acknowledgments \par}

The ARPES measurements were performed with the approval of the Proposal Assessing Committee of Hiroshima Synchrotron Radiation Center (Proposal Numbers: 17BG009, 18AG026, and 17AU026). This work was supported by the MOST of China (Grant Nos. 2016YFA0300904, 2016YFA0202301), the NSF of China (Grants Nos. 11761141013, 11674366, 11674368, 91850120), the Strategic Priority Research Program of the Chinese Academy of Sciences (Grant Nos. XDB07010200, XDB30000000), and the National Key Research and Development Program of China (Grant Nos. 2016YFA0300902 and 2015CB921001). I.M. was supported by a Grant-in-Aid for Specially Promoted Research (KAKENHI 18H03874) from the Japan Society for the Promotion of Science.

B.F., H.Z., and Y.F. contributed equally to this work.
\end{acknowledgments}

\end{document}